# MECHANISMS AND OBSERVATIONS OF CORONAL DIMMING FOR THE 2010 AUGUST 7 EVENT

James P. Mason[1], Thomas N. Woods[1], Amir Caspi[1], Barbara J. Thompson[2], and Rachel A. Hock[3]
[1] Laboratory for Atmospheric and Space Physics, University of Colorado, Boulder, CO 80303, USA
[2] Solar Physics Laboratory, NASA Goddard Space Flight Center, Greenbelt, MD 20771, USA
[3] Space Vehicles Directorate, U.S. Air Force Research Laboratory, Kirtland Air Force Base, Albuquerque, NM 87116, USA

## ABSTRACT

Coronal dimming of extreme ultraviolet (EUV) emission has the potential to be a useful forecaster of coronal mass ejections (CMEs). As emitting material leaves the corona, a temporary void is left behind which can be observed in spectral images and irradiance measurements. The velocity and mass of the CMEs should impact the character of those observations. However, other physical processes can confuse the observations. We describe these processes and the expected observational signature, with special emphasis placed on the differences. We then apply this understanding to a coronal dimming event with an associated CME that occurred on 2010 August 7. Data from the *Solar Dynamics Observatory*'s (*SDO*) Atmospheric Imaging Assembly (AIA) and EUV Variability Experiment (EVE) are used for observations of the dimming, while the *Solar and Heliospheric Observatory*'s (*SoHO*) Large Angle and Spectrometric Coronagraph (LASCO) and the *Solar Terrestrial Relations Observatory*'s (*STEREO*) COR1 and COR2 are used to obtain velocity and mass estimates for the associated CME. We develop a technique for mitigating temperature effects in coronal dimming from full-disk irradiance measurements taken by EVE. We find that for this event, nearly 100% of the dimming is due to mass loss in the corona.

## 1. INTRODUCTION

Coronal dimmings were first observed in Skylab data and characterized as transient coronal holes (Rust & Hildner 1976; Rust 1983). Subsequent similar observations (Hudson et al. 1996, 1997; Sterling & Hudson 1997) using the *Yohkoh* Soft X-ray Telescope yielded a greater understanding of the sources of the dimmings in the inner corona and their timescales.

Further studies using the *Solar and Heliospheric Observatory* (*SoHO*) Extreme-ultraviolet Imaging Telescope (EIT, Delaboudiniere et al. 1995) made clearer associations to the source of coronal mass ejections, and have established that extreme ultraviolet (EUV) dimmings are a good indicator of the apparent base of the white light CME (Thompson et al. 2000; Harrison et al. 2003; Zhukov & Auchère 2004). Thus, dimmings are usually interpreted as mass depletions due to the loss or rapid expansion of the overlying corona (Hudson et al. 1998; Harrison & Lyons 2000; Zhukov & Auchère 2004). This interpretation is supported by observations of simultaneous and co-spatial dimmings in several coronal lines (e.g., Zarro et al. 1999; Sterling et al. 2000) and spectroscopic observations (Harra & Sterling 2001).

Extended studies have begun to develop a statistical understanding of these events. Reinard & Biesecker (2008) found that coronal dimmings are more likely to occur near active regions, and typically have a rapid decrease in emission followed by a more gradual recovery, lasting from 3 to 12 hours and rarely persisting longer than one day (whereas true coronal holes tend to persist for many days). Although CMEs are also observed to occur without dimmings, Reinard & Biesecker (2009) found that non-dimming CMEs all have speeds of less than 800 km s$^{-1}$, suggesting a more intimate connection between the CME and dimming properties. Krista & Reinard (2013) found further correlations between dimming magnitudes, flares, and CME mass by studying variations between recurring eruptions and dimmings.

Similar observations can now be done with the *Solar Dynamics Observatory* (*SDO*, Pesnell et al. 2012) Atmospheric Imaging Assembly's (AIA's, Lemen et al. 2012) seven EUV channels. Additionally, dimming can now be identified in irradiance measurements such as those taken with SDO EUV Variability Experiment (EVE, Woods et al. 2012). AIA and EVE observations of coronal dimming events are the primary focus of the present paper.

Coronal dimming is of particular interest for the space weather community. CMEs, when directed toward Earth, can cause geomagnetic storms. The negative consequences of these storms on our space-based and even ground-based technology are well established (National Research Council 2008); therefore, understanding CMEs and improving predictions for CME properties are important for space weather. We aim to establish parameterizations of coronal dimming and correlate them with CME velocity and mass, two key components of CME geoeffectiveness. These coronal dimming analyses hold little promise**,** however**,** of predicting the southward component of the CME's magnetic field – the third important indicator of CME geoeffectiveness.

This paper has three main objectives: 1) to clearly identify and distinguish the various physical processes that can lead to observations that might be called dimming, 2) to characterize the dimming observed for a specific event using AIA and EVE, and 3) to develop a technique for matching EVE dimming to what is thought to be the CME-related dimming observed in AIA. This last objective is referred to as the **"EVE dimming correction method"**. The next paper planned will replicate the analysis here for multiple events during two different two-week periods in 2011, with particular emphasis on validating the dimming correction method for EVE. It will also begin the comparison of dimming parameters with CME kinetics. A motivation for this study is to relate all dimmings during the SDO mission to provide a true statistical test of the correlation between dimming parameters and CME kinetics.

The present paper will first briefly describe the instruments and observations to be used in this series, then tabulate and detail dimming physical processes and resultant observational expectations, then apply this understanding to a dimming event that occurred on 2010 August 7, and finally describe the EVE correction method for its dimming observations and show its results for this event.

## 2. INSTRUMENTS AND OBSERVATIONS

**Table 1**
**Selected emission lines in EVE; peak formation temperature from Mazzotta et al. (1998)**

| Ion | Wavelength (Å) | Peak Formation Temperature (MK) |
|---|---|---|
| Fe IX | 171 | 0.631 |
| Fe X | 177 | 0.933 |
| Fe XI | 180 | 1.15 |
| Fe XII | 195 | 1.26 |
| Fe XIII | 202 | 1.58 |
| Fe XIV | 211 | 1.86 |
| Fe XV | 284 | 2.19 |
| Fe XVI | 335 | 2.69 |
| Fe XVIII | 094 | 6.46 |
| Fe XX | 132 | 9.33 |

SDO/EVE Multiple EUV Grating Spectrograph (MEGS) A and B are two grazing-incidence, uncollimated spectrographs (i.e. no spatial resolution) with 1 Å spectral resolution covering the combined range 60 – 1060 Å. Table 1 shows the emission lines observed by EVE of most interest for this paper. Note the large number of Fe ionization states, which can be used to make inferences about solar corona plasma (e.g., Warren et al. 2013; Caspi et al. 2014). Note also that all of these emissions are from the MEGS-A channel. As long as only one flare is dominant in the time series, flare characteristics can be studied with the EVE data as discussed by (Woods et al. 2011).

SDO/AIA is a full-disk imager with seven EUV channels, each with 0.6 arcsec pixel$^{-1}$ spatial scale. The spectral resolution varies slightly between channels, but is tens of Angstroms for each channel – significantly coarser than EVE. In particular, three of the channels have centers that roughly correspond to EVE lines and are the only channels used in this study: 171 Å, 193 Å, and 304 Å.

The SOHO Large Angle and Spectrometric Coronagraph (LASCO, Brueckner et al. 1995) has two white light coronagraphs in operation. The LASCO C2 field of view extends from 2 to 6 $R_\odot$, and C3 images from 3.7 to 32 $R_\odot$. SOHO is positioned at the Earth-Sun L1 Lagrange point.

The Solar Terrestrial Relations Observatory (STEREO) is a pair of counter-rotating Sun satellites at roughly 1 astronomical unit (AU). The inner (COR1) and outer (COR2) white light coronagraphs provide images from 1.5 to 4 $R_\odot$ and out to 15 $R_\odot$, respectively. The locations of STEREO and SOHO allow the application of 3D techniques for velocity and mass determination. For this study, the Coordinated Data Analysis Workshops (CDAW) CME catalog (Gopalswamy et al. 2009) is used to identify the associated CMEs and obtain basic CME parameters.

## 3. DIMMING CLASSIFICATION

Various physical processes can lead to similar observational signals, particularly if one is focused on only a single emission line. This is particularly true in the case of EVE/MEGS-A/B (hereafter referred to as simply "EVE") data being used independently where no spatial information can be used to differentiate the source of an observed dimming. The physical processes can have similar timescales and magnitude. Fortunately, spectral information can be used to deconvolve some of the processes in order to isolate the one most relevant for CMEs, what we call mass-loss dimming. Table 2 summarizes dimming physical processes and expected observational signals and the following sections describe each in further detail.

**Table 2**
**Summary of physical processes that can manifest as observed dimming**

| Short Name | Physical Process | EVE Observational Identifiers | AIA Observational Identifiers |
|---|---|---|---|
| Mass loss (Figure 1) | Ejection of emitting plasma from corona | Simultaneous intensity decrease in multiple coronal emission lines, with percentage decrease indicative of percentage mass lost | Area over and near the erupting active region (AR) darkens |
| Thermal (Figure 2) | Heating raises ionization states (e.g., a fraction of Fe IX becomes Fe X); cooling does the opposite | Heating: Emission loss in lines with lower peak formation temperatures and near simultaneous emission gain in lines with higher peak formation temperatures; vice versa for cooling | Heating: Area near AR darkens in channels with lower peak formation temperature and near simultaneous brightening in channels with higher peak formation temperatures; vice versa for cooling |
| Obscuration (Figure 4) | Dim feature (e.g., filament material) moves into line-of-sight over a bright feature (e.g., flare arcade) | Drop of emission lines proportional to their absorption cross section in the obscuring material | Direct observation of this obscuration process |
| Wave (Figure 5) | Wave disturbance propagates globally, causing compression/rarefaction of plasma as wave passes by | No effects have been identified | Direct observation of this wave process, especially apparent with difference movies |
| Doppler Dimming | Fast moving plasma Doppler shifts away from resonant fluorescence with solar emission lines | Doppler wavelength shift of emission lines and change in intensity, possibly also observed as line broadening | Change in intensity of moving plasma as its velocity changes |
| Bandpass Shift | Emissions from fast moving plasma have Doppler wavelength shift | Emission line shifts in wavelength or has broadening | Doppler shift convolves with band-pass sensitivity to cause apparent reduction in emission |

### 3.1 Mass-loss Dimming

**Figure 1: Cartoon depicting the process of mass-loss dimming. Prior to the eruption (left), coronal loops are relatively quiescent. During and after the eruption (right), the loops are brighter and reconfigured, a CME is ejected, and a void forms in the coronal plasma.**

As the short name suggests, the physical process in mass-loss dimming is the ejection of emitting plasma (see Figure 1; Harrison & Lyons 2000; Harra & Sterling 2001). The ejection can be a CME or a failed ejection, the latter of which still manifests locally as a mass-loss dimming, but does not result in the appearance of a CME in coronagraph data. This is the physical process assumed to be the main contributor to observed dimming in many recent studies (Sterling & Hudson 1997; Reinard & Biesecker 2008, 2009; Aschwanden et al. 2009). Harrison et al. (2003) showed that dimmings can account for a large percentage of CME mass. Thus, mass-loss dimming is very relevant for the space weather community.

Observationally, mass-loss dimming appears in EVE as multiple emission lines dropping nearly simultaneously. In the case of a failed ejection, the dimming area and the ejected material are likely to maintain a total emission that is close enough to constant that it will not be apparent in EVE data. For space weather, this is of little concern since CMEs have far greater geoeffectiveness than short-lived holes in the corona of small spatial extent. However, AIA data allow the identification of mass-loss dimming even if the event is a failed ejection. In either case, mass-loss dimming appears in AIA as a relatively compact area near an AR becoming darker, sometimes with a dark cloud visibly moving off-disk. Assuming the dimmings in Reinard & Biesecker (2008) to all be due to mass loss, the timescale of the process is 3 to 12 hours and rarely persists longer than a day. Additional observations from the Hinode spacecraft have confirmed density decreases with accompanying outflows (Attril et al. 2010; Harra et al. 2011a; Tian et al. 2012).

### 3.2 Thermal Dimming

**Figure 2: Cartoon depicting the observational difference between dimming and non-dimming emission lines. Relative to a pre-eruption time (left), the Fe IX emission drops while the Fe XIV emission increases (right) due to heating of the plasma and redistribution of ionization states.**

Temperature evolution of emission lines is only interpreted as observed dimming if one is not careful to observe co-spatial emission lines at different peak formation temperatures. These "heat wave dimmings" were documented by Robbrecht & Wang (2010). As plasma is heated or cooled, the ionization fraction changes, necessarily causing the emission intensity to change (Figure 2). For example, heating causes some Fe IX to become Fe X and thus, in the absence of competing physical processes, 171 Å emission drops while 177 Å rises. This pattern was identified observationally in Figure 6 of Woods et al. (2011). It can also be observed in the standard composite (multi-wavelength) movies produced by the AIA team; indeed, this is one of the prime purposes for the composites. The initiation time and duration of temperature evolution tends to be quite similar to mass-loss dimming, as they are typically both responses to the rapid release of magnetic field energy in active regions and require several hours of recovery time. Thus, thermal processes could be mistaken for mass loss if only a single spectral line was observed. Ideally, unblended emission lines from an entire coronal ionization sequence (e.g., Fe I to Fe XVIII) could be used to mitigate this convolution of dimming observations. However, as we will show in Section 4.3, it may be sufficient to have observations of two sufficiently separated ionizations states to differentiate between thermal evolution and mass-loss dimming. This is due, in part, to the fact that hotter lines (e.g., Fe XV and above) are primarily emitted from confined loops near the flare and are thus not strongly impacted by mass-loss dimming.

It is important to note that, in general, the magnitude of total observed dimming in a given line in EVE spectra is inversely proportional to its peak formation temperature, which can be inferred from Figure 3. This figure was generated using a simple algorithm that searched the EVE catalog for relative irradiance decreases greater than a specified threshold (1%, 2%, 3%) after flares exceeding GOES X-ray class of C1. The window of time searched was bounded by the GOES event start time and the sooner of either 4 hours after the start time or the next GOES event start time. This algorithm was applied to all EVE data from mission start to 23 September 2013. Figure 3 shows that the number of dimmings dramatically decreases as the magnitude threshold is increased, and decreases slightly with higher peak formation temperature. This latter effect is due to flare heating adding emission in the higher temperature, higher ionization state, lines that partially offsets the mass-loss dimming. These trends indicate that at sufficiently high peak formation temperature, no dimming may be observed at all, even at the lowest detection threshold, which is consistent with the hotter lines being restricted to the confined flare loops and hence experiencing no mass loss.

**Figure 3: Number of identified dimmings in EVE for 6 spectral lines using different percentage dimming depths as the threshold for a detection. There were 263 flares used to trigger an automated search for dimming in EVE. Note the decrease in detections with increasing peak formation temperature.**

### 3.3 Obscuration Dimming

**Figure 4: Cartoon depicting the process of obscuration dimming. A filament previously obscuring only quiet Sun (left) expands and moves in front of a flare arcade (right). This results in a decreased observed emission from the flare arcade in wavelengths where the filament is optically thick.**

The physical process that results in apparent dimming here is material that is dark in a particular wavelength (e.g., a filament) moving between bright material (e.g., flare arcade) and the observer (Figure 4). The dark plasma absorbs some of the bright emission, resulting in an apparent decrease in emission. The slow draining of material back to the corona can obscure underlying emission for hours, and absorption can be observed in both coronal and chromospheric lines (e.g. Gilbert et al. 2013). Although the obscuration dimmings can exhibit time and spatial scales comparable to the more short-lived mass-loss dimmings, it is fairly straightforward to identify absorption signatures in the EUV images. It may also be possible to identify this phenomenon with EVE using the He II 256 Å chromospheric emission line and knowledge of the absorption cross-section through filamentary plasma. Further analysis of this type of dimming is required before any conclusions can be drawn. The event analyzed in the present paper was chosen partially for the absence of any apparent obscuration effects.

### 3.4 Wave Dimming

**Figure 5: Similar to Figure 1, but depicting the process of wave dimming. After an eruptive event, a wave propagates and expands through the corona. The compressed plasma of the wavefront results in enhanced emission, while the rarefied trailing region is dimmed.**

The debate about the physics of coronal "EUV waves" continues (e.g., Zhukov & Auchère 2004; Muhr et al. 2011; Liu 2014) but one of the simplest explanations of the observations is that plasma is compressed as a longitudinal wave passes through the medium. Traveling (i.e. not-static) rarefactions are sometimes observed following the compression (Muhr et al. 2011), the compressed regions having higher densities resulting in increased emission, and vice versa (Figure 5). The EUV waves emanating from an eruption can be seen to cause dimmings and brightenings elsewhere in the solar EUV images, often very far from the original eruption site, particularly near other active regions. We refer to these dimmings that are non-local to the erupting site as "sympathetic dimmings."

It is important to distinguish between the wave-caused dimmings and other causes of remote dimming, such as large-scale disappearing loops that are visible in soft X-ray images but only have visible EUV changes at their footpoints (Pohjolainen et al. 2005). EUV wave dimmings are unlikely to be easily identified in full-disk spatially-integrated instruments like EVE because the enhanced emission nearly cancels out the dimmed emission when summed. The event studied in this paper did have a wave, but there are no clear signals of it in EVE.

### 3.5 Doppler Dimming

Resonant fluorescence of a high-velocity, remote cloud of plasma (e.g., CME) by a source population (solar emission lines) can decrease as the resultant Doppler shift becomes sufficiently large (Hyder & Lites 1970). This phenomenon has been known for decades for cometary emissions (Swings 1941; Greenstein 1958) and has been documented in chromospheric lines associated with eruptions (Labrosse & Mcglinchey 2012) as well as in coronal lines such as O VI for polar coronal hole outflows (Giordano et al. 2000). However, the majority of emission lines in the corona are collisionally dominated and will not exhibit this effect. Therefore, it is possible to diagnose this type of dimming when it is pronounced in resonantly excited lines but does not manifest in collisionally-dominated lines.

### 3.6 Bandpass Shift Dimming

This physical process is tied to the observer's location similarly to obscuration dimming. Mass ejected toward the observer will have emissions that are necessarily Doppler-shifted in wavelength. If this shift was large enough, it could shift emission lines outside of the imager bandpass, causing an apparent dimming in the data. However, as noted in Hudson et al. (2011), these Doppler shifts tend to be on the order of picometers while the bandpasses of EUV imagers are on the order of nanometers. Thus, this type of apparent dimming is not expected in EUV images, but we mention it in this paper for completeness.

In a spectrograph like EVE, the Doppler shifts would instead simply cause a wavelength shift of the emission line from the ejected material. When this Doppler-shifted emission is added with the relatively static plasma remaining on the Sun, a small Doppler shift from the ejected material would manifest as line-broadening in the integrated irradiance while a large shift would result in a line splitting. It should be noted that the EVE extracted lines data product applies a static mask to the spectra so a sufficiently large Doppler shift could cause an apparent dimming in this product. Again, the observed shifts are far too small to impact the EVE data analysis.

## 4. ANALYSIS OF 2010 AUGUST 7 EVENT

### 4.1 Coronagraph Observations

**Figure 6: CME event at 19:00 on 2010 August 7. (Left) Difference image from LASCO C2 and AIA 193 Å channel. (Right) CME height versus time shows nearly linear velocity of 871 km s$^{-1}$. (Figure adapted from CDAW CME Catalog, courtesy of S. Yashiro and N. Gopalswamy)**

The CDAW catalog has seven CME events listed for 2010 August 7. All but two of them occur prior to the M1.0 flare at 18:24 UT that is of primary interest for the present study. The CME shown in Figure 6 is flagged as a halo event with a time of 18:36 UT in CDAW, while the next event occurred with a central position angle of 116° at 22:24 UT. The timing and location of the flare and associated dimming region suggest that the halo CME is the mass associated with the dimming. The plane-of-sky velocity estimate for this CME is 871 km s$^{-1}$ as indicated in Figure 6. No mass is listed for this CME in CDAW, but using LASCO and STEREO data and the techniques outlined in Colaninno & Vourlidas (2009), a mass of 6.4 x10$^{15}$ g was computed for this CME event (Vourlidas **2014**, private communication). A true space velocity was also computed as 850 km s$^{-1}$ at 9 R$_\odot$ with a deceleration of 6.84 m s$^{-2}$ (Figure 7). Based on these estimates for mass and velocity, this CME is considered be of modest size.

**Figure 7: (Left) STEREO-A COR2 image at 19:24UT. (Right) CME height versus time calculated from STEREO and shows a deceleration of 6.84 m s$^{-2}$.**

### 4.2 AIA EUV Image Observations

**Figure 8: AIA Results for the M1.0 Flare on 2010 August 7. (Top) AIA 171 Å channel difference image with subjectively selected region contours overlaid. The red contour outlines what is thought to be the region of mass loss. The orange and purple contours outline other active regions on the disk, which have the potential to have sympathetic dimming. The green contour outlines a filament, which also has the potential to sympathetically dim based on its behavior during the M flare on 5 August. The magenta contour isolates the flaring coronal loops. The white line around the solar limb is an artifact of the solarsoft derotation method. (Bottom three plots) Light curves of AIA 171 Å, 193 Å, and 304 Å channels for the color-corresponding contours on the AIA image. The blue line is the light curve for all on-disk area not enclosed by a contour. The black line is the sum of all contoured regions and acts as a proxy for total dimming. All percent changes are calculated from the band's value at 17:00 UT, prior to the flare. The transition**

**region He emission does not show dimming; both corona Fe emissions show dimming.**

The relative simplicity of this event is why it was chosen for a case study. The observations in AIA do not suggest that obscuration, waves, or Doppler shift contributed to the observed dimming. The area in the red contour of Figure 8 was selected manually (by eye) to represent the region of mass loss, and its light curve shows clear dimming in 193 Å and 171 Å. In fact, the dimming from this region accounts for nearly all of the observed dimming throughout the entire event. This contour was selected after several iterations that indicated slight deviations in the contour had minimal impact on the light curve, as long as the dark region was fully encompassed. The other contours were also selected manually to isolate regions that appear to dim. The exception is the magenta contour surrounding the flare loops that brightens dramatically but doesn't ever dim.

The He II 304 Å light curves are included to provide a contrast to the dimming effects seen in the coronal Fe lines. This He II wavelength is generated primarily in the chromosphere and transition region, as opposed to the coronal source of the other wavelengths. Mass loss occurs primarily in the corona, as the term coronal mass ejection suggests. This is reflected in the lack of dimming observed in the non-coronal He II 304 Å line.

Thermal dimming may play a role in this event but may be difficult to quantify using only AIA because the relatively wide spectral bands of AIA channels mean many emission lines and any continua are blended together, which makes specifying a well-defined temperature difficult. EVE is less sensitive to this issue due to its higher spectral resolution and plethora of emission lines from Fe at different ionization states. A future study using the differential emission measure techniques of Caspi et al. (2014) to study the temperature evolution could help to quantify this effect.

### 4.3 EVE EUV Irradiance Observations

**Figure 9:** One minute average EVE light curves of the 2010 August 7 coronal dimming event for the spectral lines listed in Table 1, as well as the GOES 1-8 Å channel light curve. The leftmost vertical dashed line indicates the GOES event start time, while the other vertical dashed line indicates the GOES event peak time. Peak formation temperature of the EVE spectral lines increases from top to bottom plot. Fe IX to Fe XIII show clear dimming, Fe XIV is borderline, and Fe XV to Fe XX show smooth brightening with no dimming. The Fe XX 131 Å profile is very similar to GOES 1-8 Å, indicating that this line in EVE is a good proxy for gradual phase timing. Also note the vertical axes: dimming is on the order of a few percent for the cooler Fe emissions while the hotter Fe emissions have bright peaks in the hundreds of percent. All percent changes are calculated from the spectral irradiance at 17:00 UT.

Figure 9 shows a trend that is consistent with the findings from Figure 3 – that an ion's peak formation temperature is inversely proportional to magnitude of dimming. The transition from a line that shows dimming to ones that only show brightening occurs at Fe XIV 211 Å, which itself shows dimming in some events but not others. The transition for where the Fe emission shows dimming is different for different flares. For example, the Fe XVI 335 Å emission has shown dimming for larger CME events (Woods et al. 2011). Analysis exemplifying these cases will be presented in a future paper. Herein, we will refer to Fe IX 171 Å through Fe XIV 211 Å as dimming lines for the 2010 August 7 event, and Fe XIV 211 Å through Fe XXIV 192 Å as non-dimming lines (note that 211 Å is included in both descriptions to reflect its ambiguity).

### 4.4 Discussion and Comparison of Observations

The dimming lines in EVE show that dimming began at the flare onset and before the peak of any of the Fe emissions. This dimming start time at about 18:05 UT is most clear in the Fe IX 171 Å emission, following a small peak that may be associated with the flare's impulsive phase. Then a larger peak is seen in the Fe IX and other Fe emissions at about 18:20 UT after the dimming had started. This larger peak is associated with the flare's gradual phase and is consistent with post-flare cooling because the peak time for the cooler Fe emissions occurs later than the hotter Fe emissions. The AIA observations allow us to easily isolate the brightening due to the flare core region (Figure 8, magenta contour) from the mass loss dimming (Figure 8, red contour). When those two light curves get convolved as they are in EVE observations, it results in the flare peak coinciding with the early dimming profile. Since this peak and any longer duration irradiance enhancements can impact parameterizations of the dimming (e.g., slope and depth) that would be used for correlation with CME kinetics, it is desirable to develop a method to mitigate its impacts as discussed later.

The non-dimming lines have a large gradual phase peak but no dimming (Figure 9). Some of them also show a slight post-flare brightening enhancement to the irradiance. To summarize the processes, the plasma and its irradiance have source and sink terms. Near the beginning of the flare, heating is very dominant and causes a rapid increase in high ionization states for the various Fe emissions. Later in the flare, cooling of the plasma causes an increase in lower ionization states, and those cooler lines peak later than the hot lines. Through it all, the mass ejection can act as a sink for most coronal emissions. Early in the flare, before the low ionization states have been strongly affected by the cooling described above, the mass ejection dominates and causes the irradiance to visibly drop. Much later in the flare process, as the plasma approaches its pre-flare level, the missing plasma again becomes apparent in the irradiance time series as an hours-long, few-percent decrease.

The expectation for mass-loss dimming is that the amount of dimming is proportional to the mass loss and that all corona Fe emissions originating in the CME initiation region (i.e., not the confined, flaring loops) would have the same level of dimming. It is obvious from the EVE time series for this event (and many other events) that the dimming amount decreases with the hotter Fe emissions. While thermal dimming could play a role in such behavior, it is known from the AIA analysis that thermal dimming is not an important contribution for this 7 August event. In particular, the amount of AIA dimming of the red traces in Figure 8 is about 3% for both the 171 Å and 193 Å bands. In contrast, the amount of dimming relative to the level near 18:05 UT is about 3% and 2%, respectively for the EVE measurements of Fe IX 171 Å and Fe XII 195 Å. Because EVE includes the full-disk irradiance, we expect that the EVE data need to be corrected for the gradual phase peak effects in the time series to isolate better the mass-loss dimming contribution.

Our method to remove the gradual phase peak in the EVE dimming time series is simple: temporally align the peak in a non-dimming line with that in a dimming line, scale the non-dimming peak down to match the dimming line peak intensity, and subtract this renormalized non-dimming time series from the dimming time series. Since we deal solely in percentage changes relative to the pre-flare irradiance, this method does not imply a direct irradiance alteration. For this event, we ran every combination of dimming and non-dimming line to determine which correction of the gradual phase peak could best match the EVE dimming result with the AIA dimming result. As a first approximation, we compare the resultant "corrected" EVE light curve to the mass-loss dimming identified in AIA (Figure 8). The best performing correction is shown in Figure 10. The correction method significantly reduces the impact of the flare's gradual phase peak to dimming measurements for EVE. Prior to the correction, EVE would have measured a dimming depth of 1.27% in 171 Å and 0.18% in 195 Å. After the correction, these values are 2.94% and 2.09%, respectively. Similarly, slope was changed from 1% $hr^{-1}$ (171 Å) and 0% $hr^{-1}$ (195 Å) to 2.29% $hr^{-1}$ (171 Å corrected) and 2.09% $hr^{-1}$ (195 Å corrected). Furthermore, if this event was being observed in real time, the gradual phase peak makes it impossible to estimate the amount and speed (slope) of dimming accurately. This correction method allows the irradiance increase due to the gradual phase contribution to be compensated in the EVE time series that have dimming.

The small difference in time between different emission peaks -- Fe XX peaks 21 minutes before Fe IX in this case -- is information that can be used to understand the temperature evolution during dimming. In this event, that time difference is significantly shorter than the hours-long duration of the total dimming event. Thus, it is unlikely that thermal dimming is a significant contributor to the total observed dimming. Instead, our correction method uses non-dimming lines as independent measurements of the flare gradual phase profile. Since no dimming is observed in the non-dimming lines, the gradual phase profile is assumed to be pure and can then be used as a proxy to remove only the effect of the gradual phase in the dimming light curve with a minimal impact on total dimming. In this way, we can effectively match AIA dimming observations, which are capable of isolating the flaring coronal loops.

The expectation is that the EVE-corrected dimming results should have the same amount of dimming as the AIA results and are also independent of Fe ionization level (in the dimming lines). Figure 10 shows the comparison of EVE-corrected dimming time series to AIA results, and Table 3 lists the dimming results. Fe XV 284 Å is best for correcting the EVE Fe IX 171 Å and Fe XII 195Å dimming to match the AIA 171 Å and 193 Å dimming amounts, respectively. Figure 9 shows that Fe XV 284 Å has a relatively smooth peak and its irradiance remains positive for many hours after the flare. This behavior is needed in the corrected spectral line in order to match the AIA observations.

Dimming is parameterized using the percent depth and slope (see Table 3). The time to use in these parameterizations is debatable, but here we chose the point where AIA region 1 dimming leveled out. The overlapping vertical red and blue arrows in Figure 10 indicate this point. The start point for the calculation of slope was chosen as 17:50 UT – the time just before GOES 1 – 8 Å and EVE 131 Å began to rise (see Figure 9).

AIA Region 1 is considered the reference for mass-loss dimming, so its dimming depth and slope are compared as an estimate of uncertainty for these results from EVE. The differences for the AIA 171 Å and 195 Å dimming depth and slope are 0.3% and 0.16% $hr^{-1}$, respectively. The relative uncertainty of these is 10% of the mean depth and slope values, being 3.02% and 1.60% $hr^{-1}$. These differences in the two different AIA bands could reflect the uncertainty that Region 1 is only due to mass-loss dimming and our ability to identify the best Region 1 boundary to encompass the mass-loss dimming phenomena. The corrected EVE results for dimming depth and slope have mean values of 2.53% and 1.34% $hr^{-1}$, and both are 14% less than the AIA Region 1 mean values. The standard deviations for the six EVE lines' corrected dimming depth and slope are 0.21% and 0.11% $hr^{-1}$, respectively. As expected (intended), the EVE corrected results are much more self-consistent with each other than the uncorrected results. The slope tracks the depth variation well; that is, the slope is less when the depth is less. Therefore, the slope result may not be providing any useful information beyond the depth result. Our expectation was that the slope could represent the CME velocity, and the depth could represent the CME mass loss. Analysis of more flare events with large dimmings is needed to address this assumption, and such analysis will be the focus of our next paper on this topic.

**Figure 10: Comparisons of EVE and AIA light curves. The EVE corrections that make the light curves most closely match AIA mass-loss dimming are shown. The vertical arrows indicate the time where slope and depth are calculated from 17:00 UT.**

5. CONCLUSION AND FUTURE WORK

Table 3 summarizes the key results found in this study and establishes the template to be used in future studies of more events. The table includes data for all EVE dimming lines, even when no corresponding AIA channel is available. In fact, many of these lines are blended together in AIA channels. For example, the 171 Å channel in AIA includes contributions from the 177 Å and 180 Å lines and the 193 Å channel includes contributions from the 180 Å, 202 Å, and 211 Å lines.

In cases where obscuration dimming is not a concern, it may be possible to isolate mass-loss dimming using only spectral information. The method presented here for removing flare induced irradiance from EVE dimming lines shows good agreement with AIA light curves that have extracted the flare coronal loops. Future work will apply the EVE correction technique and AIA validation to multiple events in order to refine the methods. That study will enable a preliminary correlation to be derived between mass-loss dimming and CME kinetic parameters. Following that, a full statistical study of the SDO period will establish the strength of the correlation between these parameters and has potential to establish relationships to predict CME parameters from the coronal dimming observations.

**Table 3**
**Key results for EVE and AIA. The associated CME had a velocity of 871 km s$^{-1}$ and mass of 6.4 x10$^{15}$ g. Empty cells indicate wavelengths that are not distinguished or available in AIA. Fe XV 284 Å is used for the EVE correction as described in Section 4.4. See Figure 10 for the plot corresponding to much of the information in this table.**

| Dimm line (Å) | AIA Total Depth (%) | AIA Reg. 1 Depth (%) | EVE Depth Corrected (%) | EVE Depth Uncorrected (%) | AIA Total Slope (% hr$^{-1}$) | AIA Reg. 1 Slope (% hr$^{-1}$) | EVE Slope Corrected (% hr$^{-1}$) | EVE Slope Uncorrected (% hr$^{-1}$) |
|---|---|---|---|---|---|---|---|---|
| 171 | 2.03 | 3.17 | 2.60 | 1.63 | 1.07 | 1.68 | 1.38 | 0.86 |
| 177 | … | … | 2.79 | 1.89 | … | … | 1.48 | 1.00 |
| 180 | … | … | 2.87 | 1.98 | … | … | 1.52 | 1.05 |
| 195 | 1.68 | 2.87 | 2.46 | 1.52 | 0.89 | 1.52 | 1.30 | 0.81 |
| 202 | … | … | 2.31 | 1.60 | … | … | 1.22 | 0.85 |
| 211 | 0.52 | 2.03 | 2.57 | 1.60 | 0.28 | 1.50 | 1.36 | 0.85 |

## ACKNOWLEDGEMENTS

The authors would like to thank Angelos Vourlidas for providing computations of mass and true velocity for the event discussed in this paper. Additionally, the CDAW CME catalog is generated and maintained at the CDAW Data Center by NASA and The Catholic University of America in cooperation with the Naval Research Laboratory. SOHO is a project of international cooperation between ESA and NASA. This research is supported by the NASA SDO project, NASA grant NAS5-02140.

FIGURES:

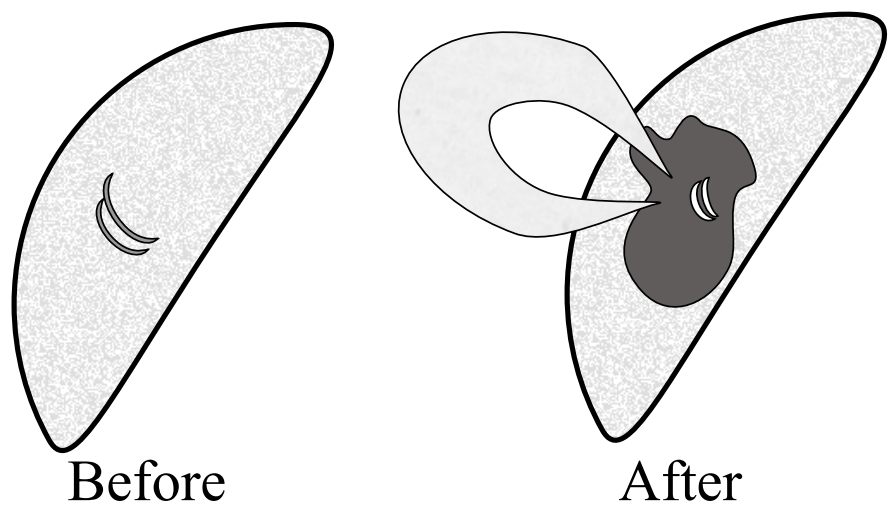

**Figure 1** – Cartoon depicting the process of mass-loss dimming. Prior to the eruption (left), coronal loops are relatively quiescent. During and after the eruption (right), the loops are brighter and reconfigured, a CME is ejected, and a void forms in the coronal plasma.

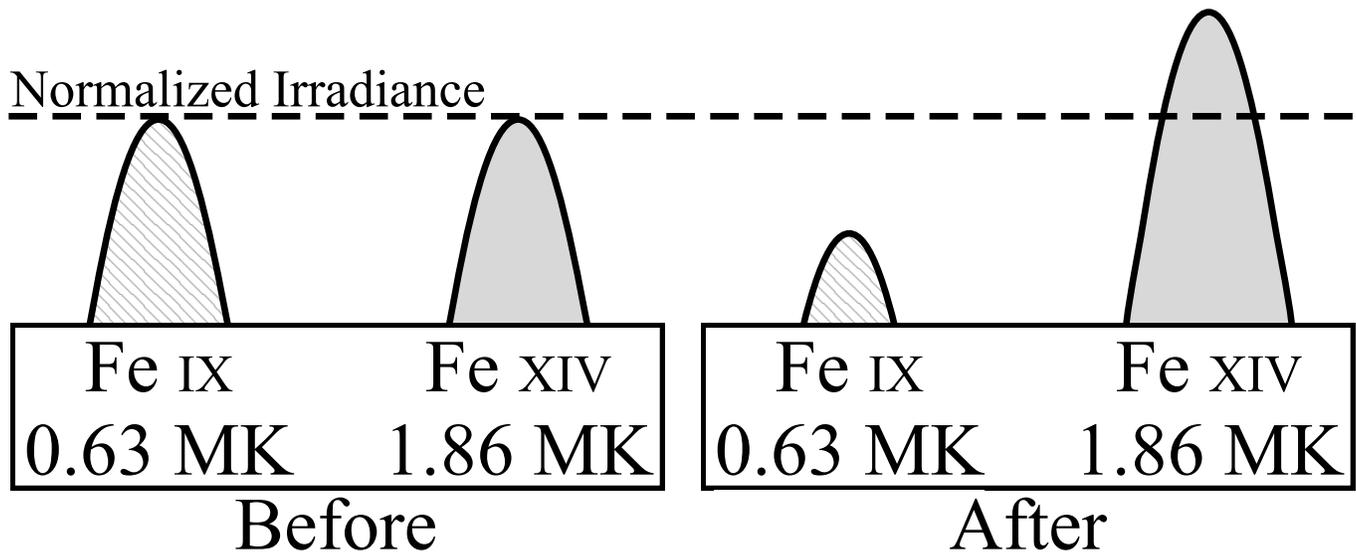

**Figure 2** – Cartoon depicting the observational difference between dimming and non-dimming emission lines. Relative to a pre-eruption time (left), the Fe IX emission drops while the Fe XIV emission increases (right) due to heating of the plasma and redistribution of ionization states.

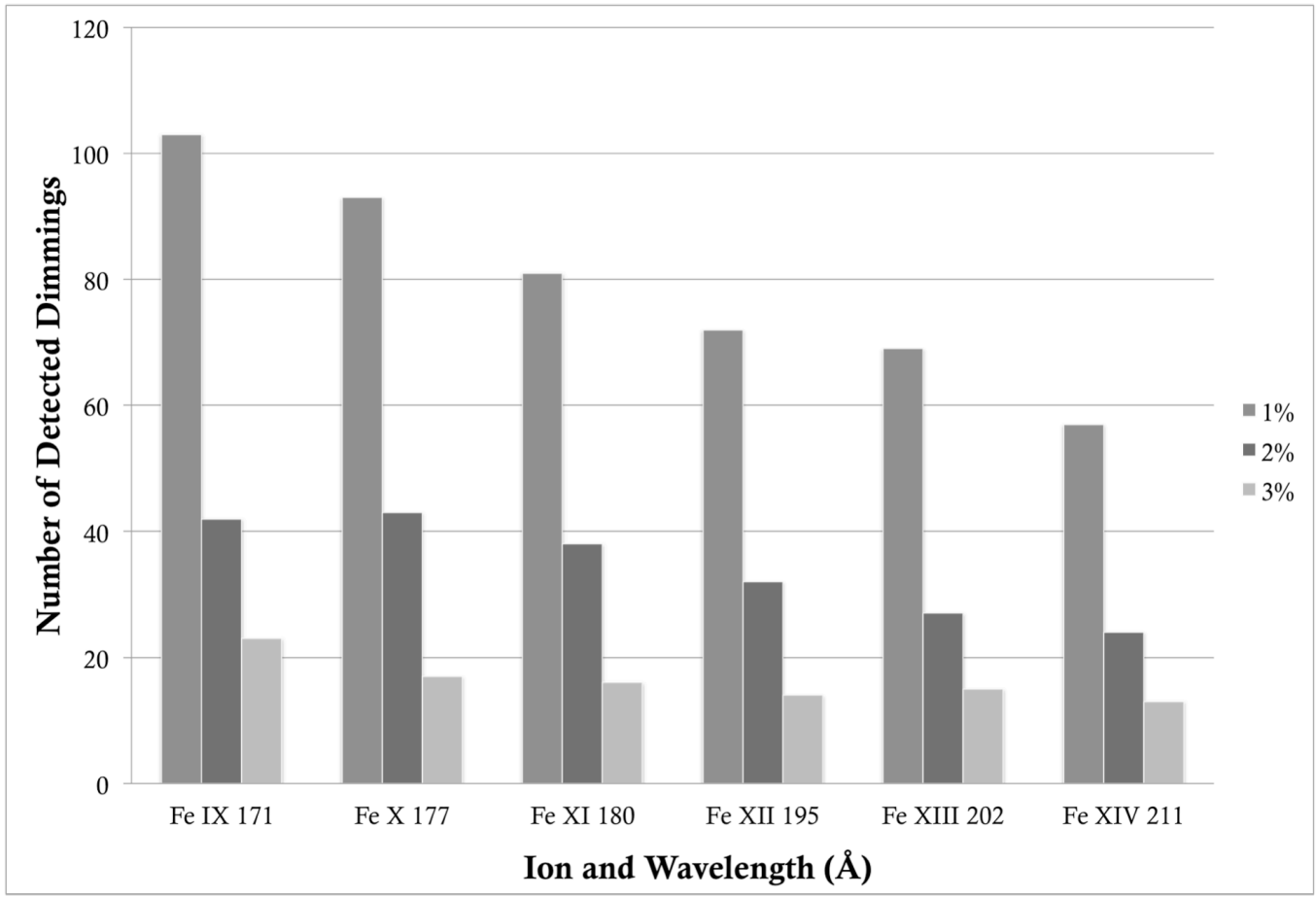

**Figure 3** – Number of identified dimmings in EVE for 6 spectral lines using different percentage dimming depths as the threshold for a detection. There were 263 flares used to trigger an automated search for dimming in EVE. Note the decrease in detections with increasing peak formation temperature.

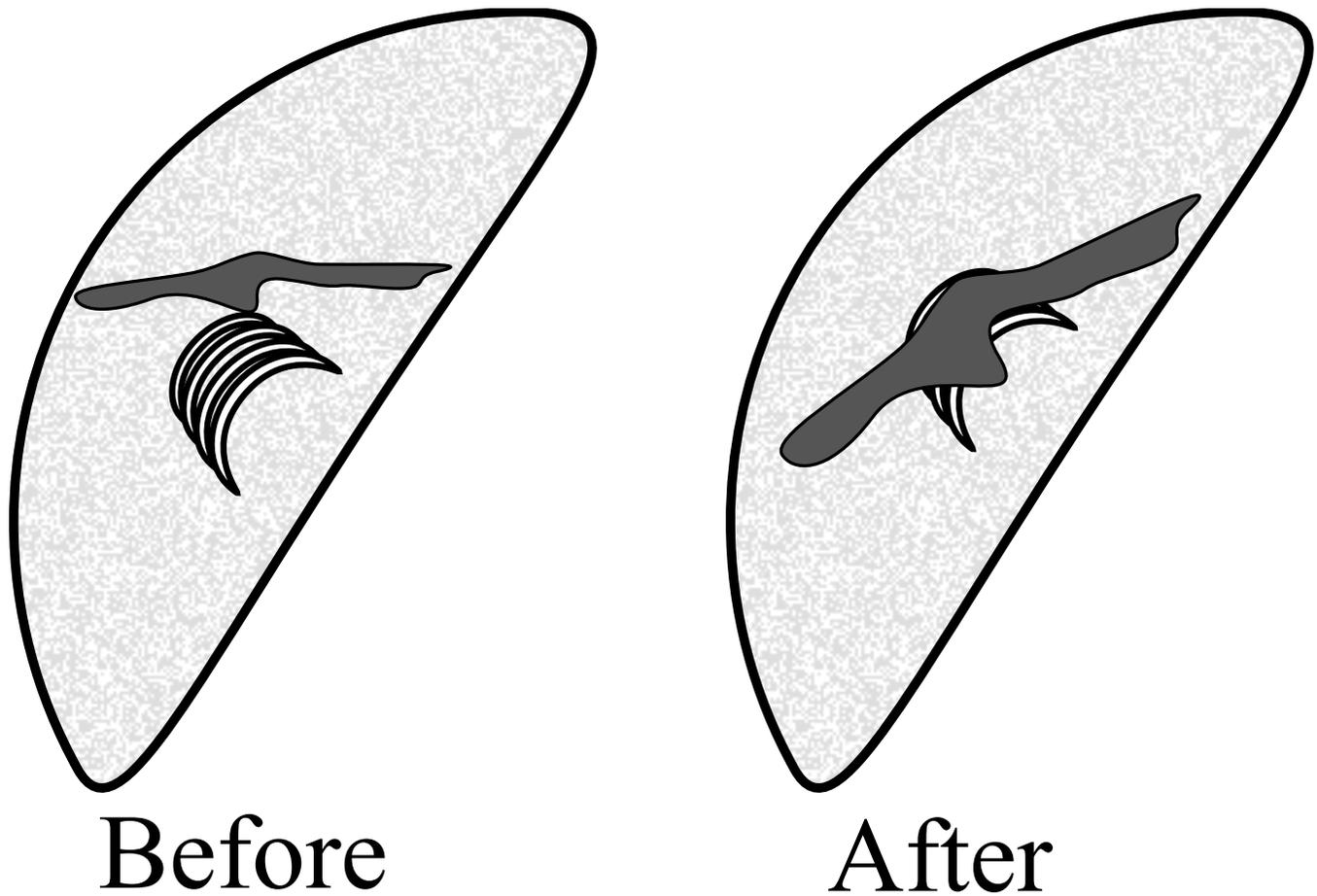

**Figure 4** – Cartoon depicting the process of obscuration dimming. A filament previously obscuring only quiet Sun (left) expands and moves in front of a flare arcade (right). This results in a decreased observed emission from the flare arcade in wavelengths where the filament is optically thick.

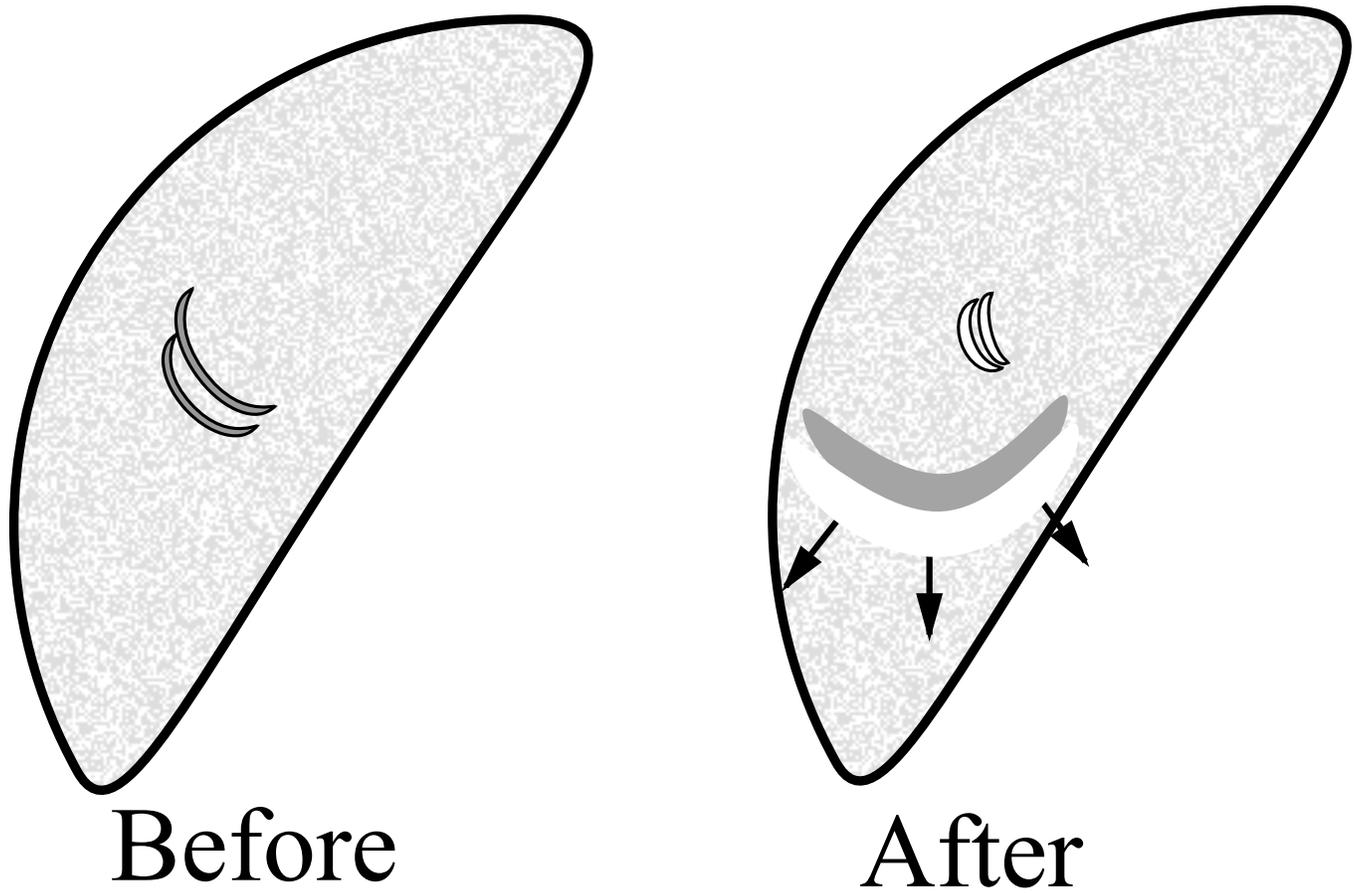

**Figure 5** – Similar to Figure 1, but depicting the process of wave dimming. After an eruptive event, a wave propagates and expands through the corona. The compressed plasma of the wavefront results in enhanced emission, while the rarefied trailing region is dimmed.

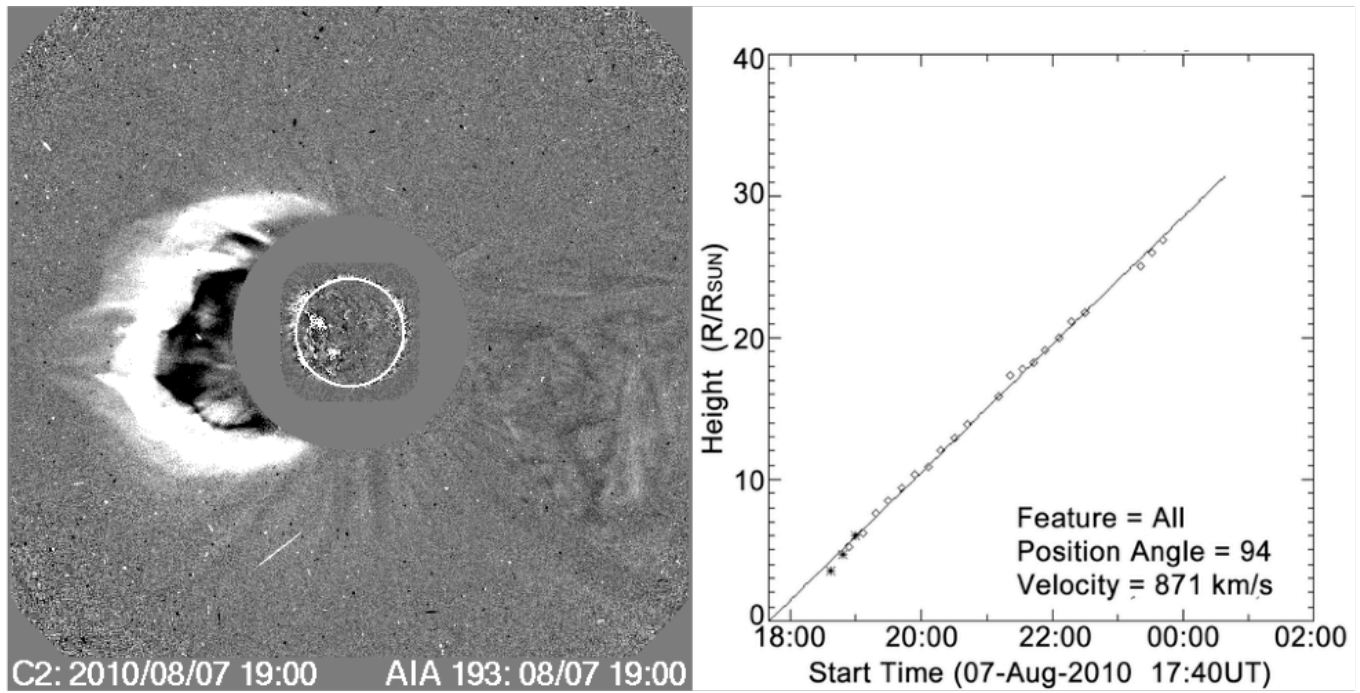

**Figure 6** – CME event at 19:00 on 2010 August 7. (Left) Difference image from LASCO C2 and AIA 193 Å channel. (Right) CME height versus time shows nearly linear velocity of 871 km s$^{-1}$. (Figure adapted from CDAW CME Catalog, courtesy of S. Yashiro and N. Gopalswamy)

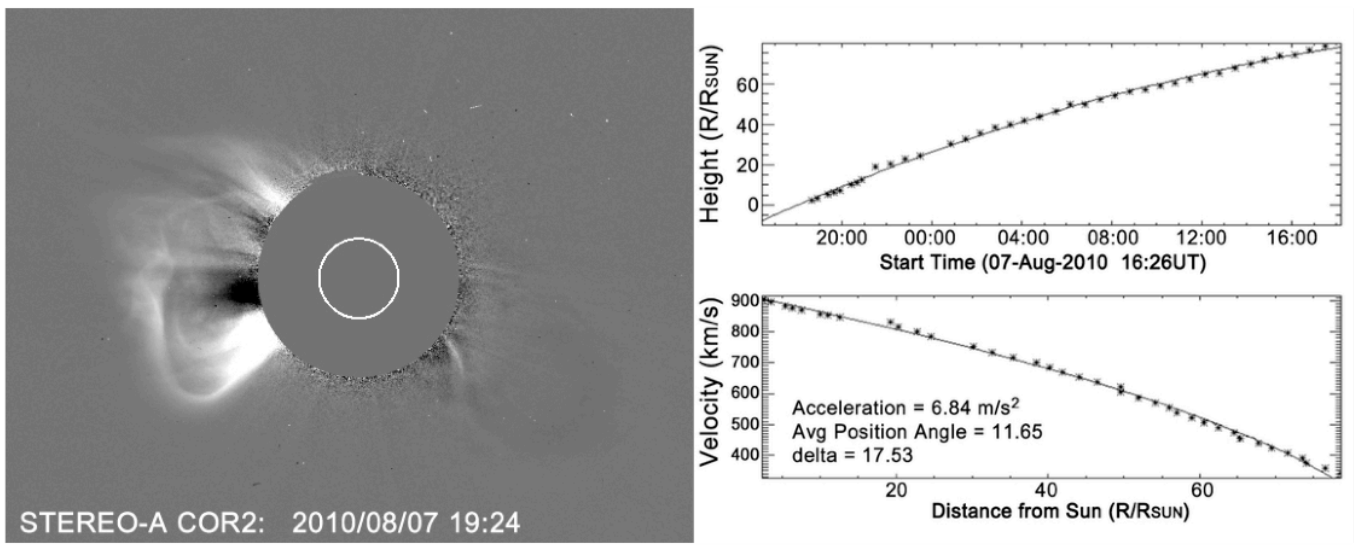

**Figure 7** – (Left) STEREO-A COR2 image at 19:24UT. (Right) CME height versus time calculated from STEREO and shows a deceleration of 6.84 m s$^{-2}$.

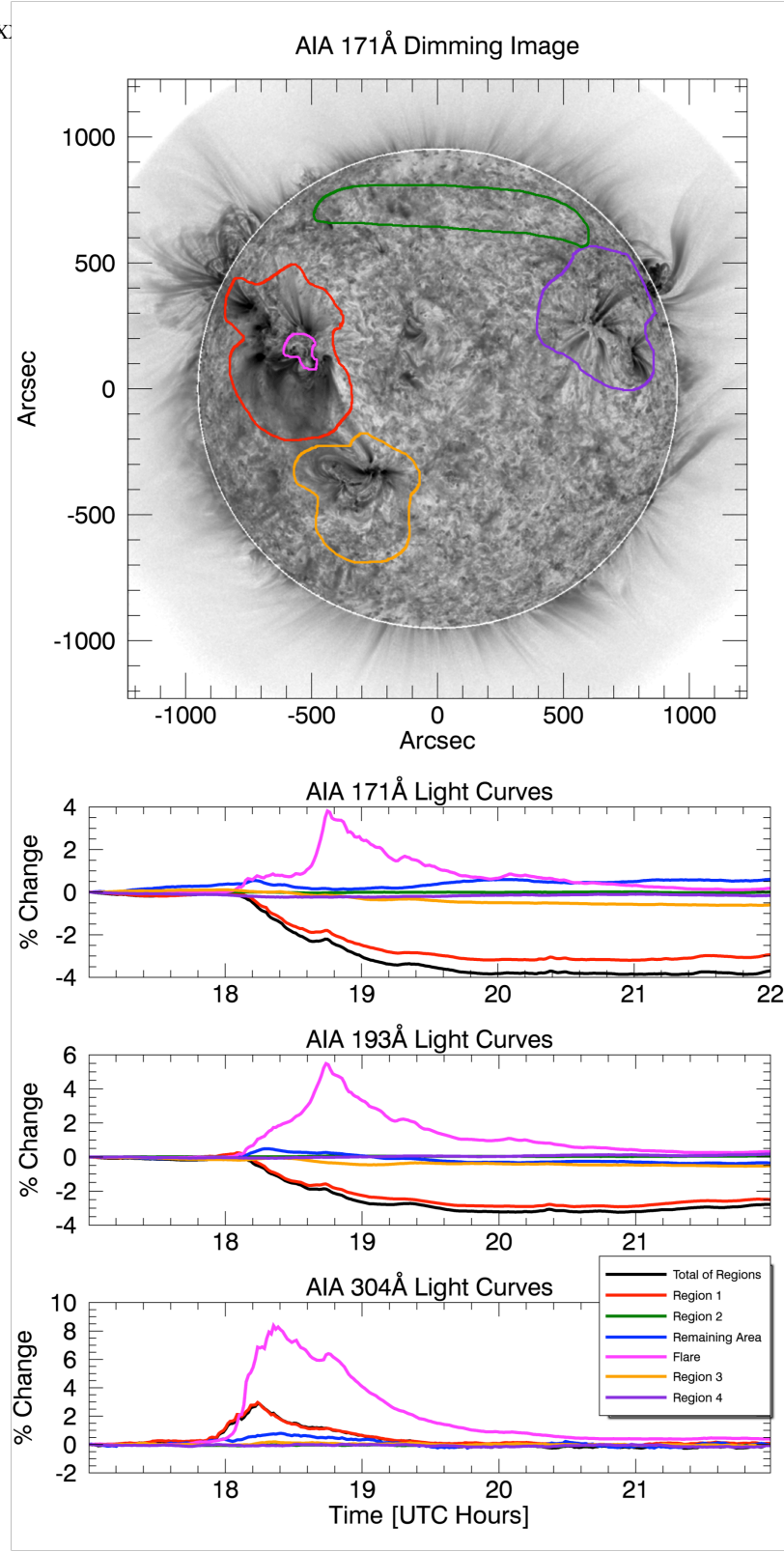

**Figure 8** – AIA Results for the M1.0 Flare on 2010 August 7. (Top) AIA 171 Å channel difference image with subjectively selected region contours overlaid. The red contour outlines what is thought to be the region of mass loss. The orange and purple contours outline other active regions on the disk, which have the potential to have sympathetic dimming. The green contour outlines a filament, which also has the potential to sympathetically dim based on its behavior during the M flare on 5 August. The magenta contour isolates the flaring coronal loops. The white line around the solar limb is an artifact of the solarsoft derotation method. (Bottom three plots) Light curves of AIA 171 Å, 193 Å, and 304 Å channels for the color-corresponding contours on the AIA image. The blue line is the light curve for all on-disk area not enclosed by a contour. The black line is the sum of all contoured regions and acts as a proxy for total dimming. All percent changes are calculated from the band's value at 17:00 UT, prior to the flare. The transition region He emission does not show dimming; both corona Fe emissions show dimming.

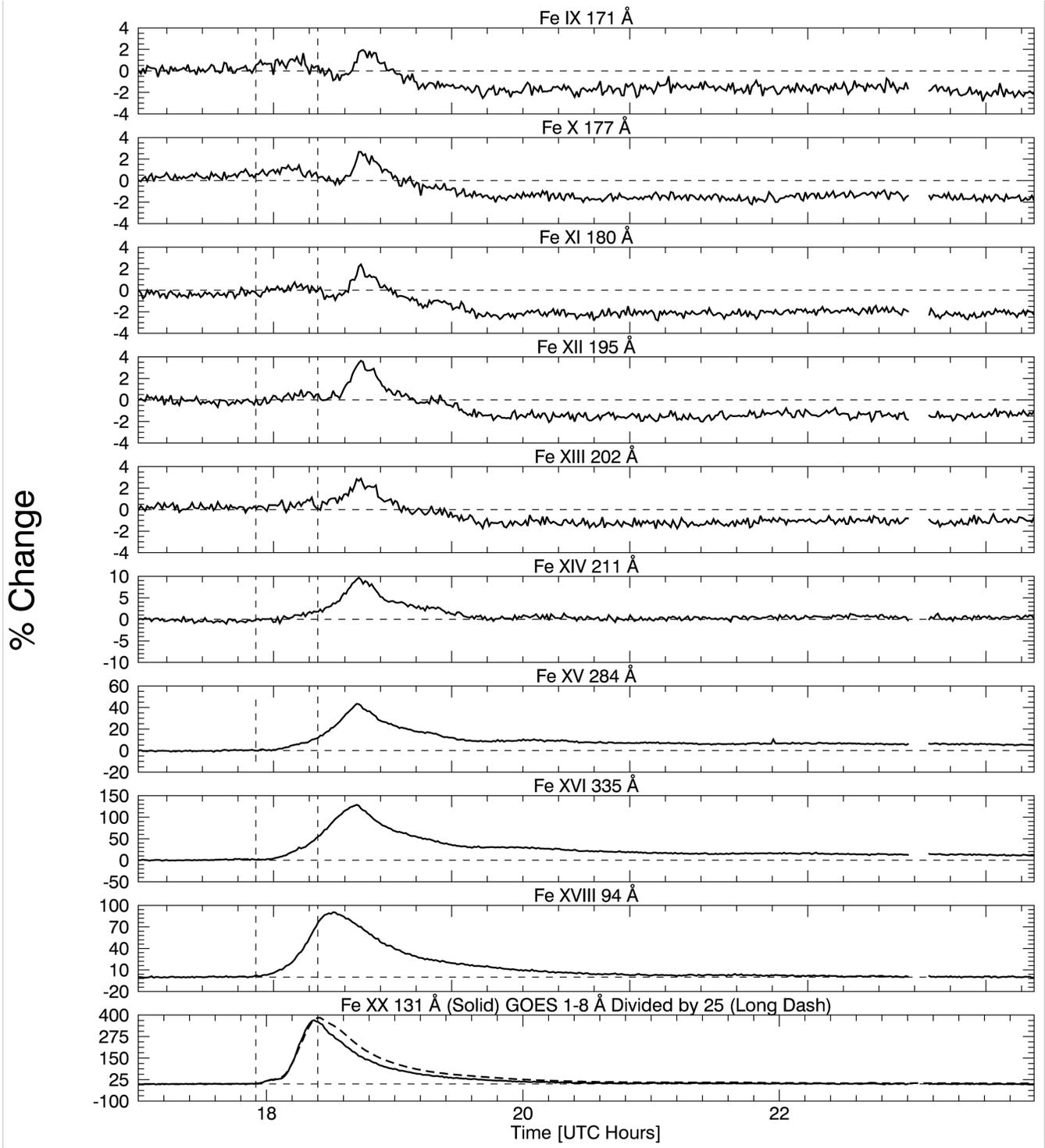

**Figure 9** – One minute average EVE light curves of the 2010 August 7 coronal dimming event for the spectral lines listed in Table 1, as well as the GOES 1-8 Å channel light curve. The leftmost vertical dashed line indicates the GOES event start time, while the other vertical dashed line indicates the GOES event peak time. Peak formation temperature of the EVE spectral lines increases from top to bottom plot. Fe IX to Fe XIII show clear dimming, Fe XIV is borderline, and Fe XV to Fe XX show smooth brightening with no dimming. The Fe XX 131 Å profile is very similar to GOES 1-8 Å, indicating that this line in EVE is a good proxy for gradual phase timing. Also note the vertical axes: dimming is on the order of a few percent for the cooler Fe emissions while the hotter Fe emissions have bright peaks in the hundreds of percent. All percent changes are calculated from the spectral irradiance at 17:00 UT.

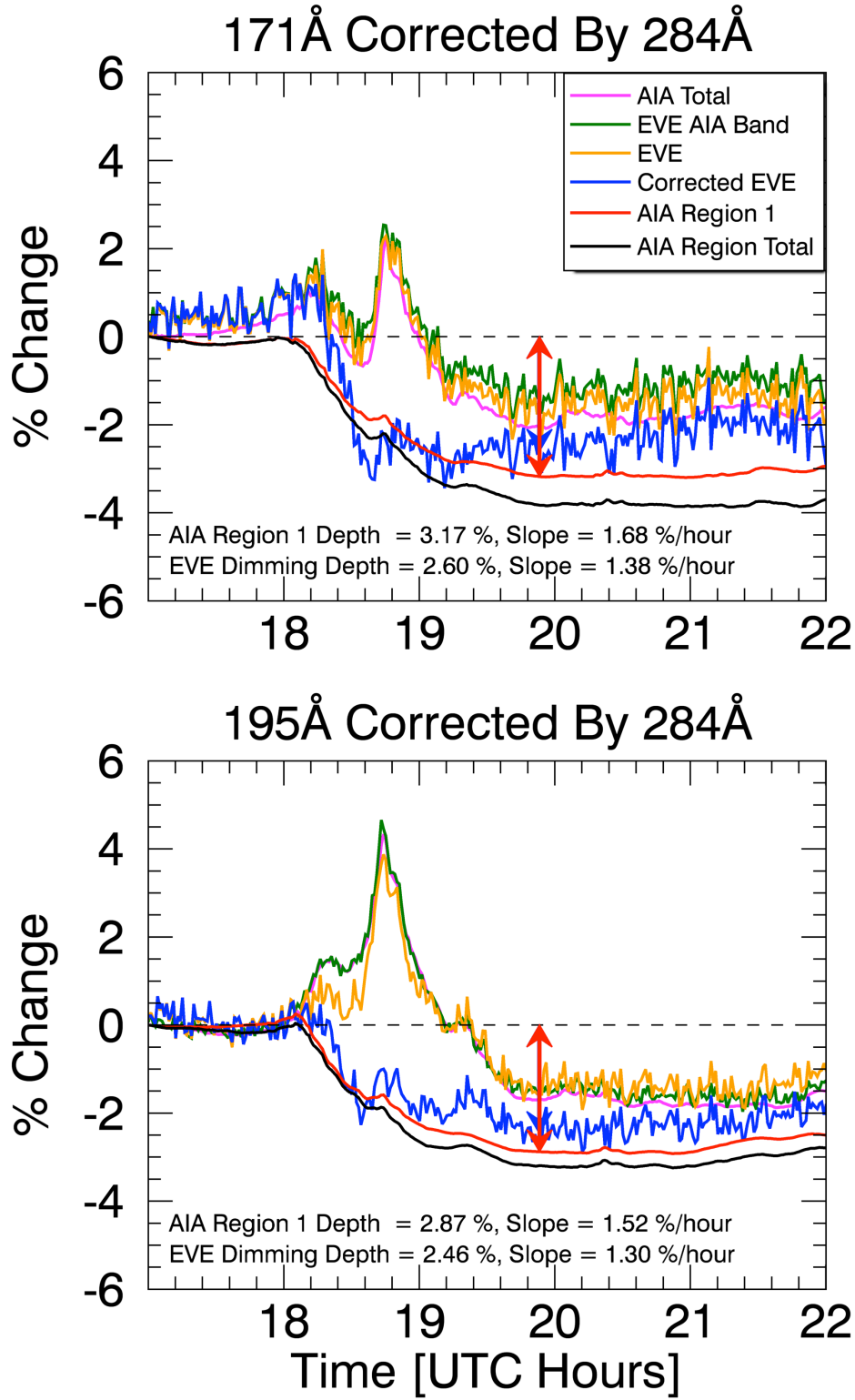

**Figure 10** – Comparisons of EVE and AIA light curves. The EVE corrections that make the light curves most closely match AIA mass-loss dimming are shown. The vertical arrows indicate the time where slope and depth are calculated from 17:00 UT.